\documentclass[twocolumn,showpacs,preprintnumbers,amsmath,amssymb,showkeys,nofootinbib]{revtex4}

\usepackage{graphicx}
\usepackage{dcolumn}
\usepackage{bm}

\begin{document}

\preprint{APS/123-QED}

\title{Bimodality as a signal of Liquid-Gas phase transition in nuclei?}

\author{O. Lopez}
 \email{lopezo@lpccaen.in2p3.fr}
\author{D. Lacroix}
\author{E. Vient}

\affiliation{%
Laboratoire de Physique Corpusculaire \\IN2P3-CNRS/Universit\'e de Caen/ENSICAEN\\
6, Boulevard Mar\'echal Juin, F-14050 Caen, France}%

\date{\today}

\begin{abstract}

We use the HIPSE (Heavy-Ion Phase-Space Exploration) 
Model to discuss the origin of the bimodality in charge asymmetry observed 
in nuclear reactions around the Fermi energy. 
We show that it may be
related to the important angular momentum (spin) transferred 
into the quasi-projectile before secondary decay. As the spin overcomes the critical value, a sudden 
opening of decay channels is induced and leads to a bimodal distribution for
the charge asymmetry. In the model, it is not assigned 
to a liquid-gas phase transition but to specific instabilities 
in nuclei with high spin. Therefore, we propose to use these reactions to study instabilities in rotating 
nuclear droplets. 
\end{abstract}


\maketitle
In recent years, the possibility to observe phase transitions in finite, even small 
systems, has received an increasing interest \cite{Gro01,Ric01,Cho04,Cho05}. In this context, 
several conceptual questions 
are addressed, as the 
extrapolation of thermodynamical properties, or more generally Statistical Mechanics, 
from finite to infinite systems. For instance, a system 
with phase transition has discontinuity in the equation of 
state in the transition region. Such a discontinuity is not present in finite 
systems but is expected to be replaced by anomalies in specific statistical 
quantities.
One of the possible signature of liquid-gas phase transition in finite systems is the appearance 
of abnormal fluctuations of the kinetic energy \cite{Cho99} in the microcanonical ensemble, 
these being at the origin of the so-called 
negative heat capacity. Equivalently, a bimodal behaviour, i.e. 
two "bumps" in the energy distribution 
is expected in the coexistence region, the system being treated canonically.    
Bimodalities in event distributions is even sometimes promoted as 
one of the definition of phase transition in finite systems \cite{Cho05,Cho03}. 
Indeed, it is related to the 
anomalous curvature of the
entropy or any relevant thermodynamical potential depending on the constraints 
upon the system (see however \cite{Tou05}). 
Moreover, several studies have shown that bimodality is rather robust with 
respect to the introduction of additional constraints \cite{Gul04} 
on the system or the long-range coulomb force \cite{Gul03}. In this context, 
nuclei appear as possible candidates to observe liquid-gas phase transition
in finite quantum system. Indeed, a large variety of experimental studies
\cite{Riv05,Dag05,Dag00,Ell01,Tab02,Fra00,Tsa01,Tra00,Wad04,Nat02, Ma05, Tam03, Pic05} 
reports an accumulation of "evidences" of critical signals. 
Among them, bimodality in the charge asymmetry of fragments produced in heavy-ion 
reactions at Fermi energies 
have been recently reported for the
Quasi-Projectiles (QP) isolated in peripheral reactions \cite{Tam03,Pic05}. 
This signal is presented as one of the most robust evidence 
for the liquid-gas phase transition in nuclei.
However, the extraction of critical signals from 
nuclear reactions is far from being simple. 
For example, one expects that a signal initially present at the chemical 
freeze-out (i.e. when nuclei do not exchange particles anymore) 
will be largely distorted, or even completely washed out by the secondary decay.
This raises the fundamental question of the phase-space explored during the reaction
just after fragment formation and its modification due to secondary emission.        
Recently, the phenomenological HIPSE\footnote{available at http://lpccaen.in2p3.fr/theorie/theory\_lacroix.html}  
model \cite{Lac04} has been developed to address these aspects.
In this model, very specific randomness hypotheses are retained to form 
clusters in the first instants of the reaction, while information on
the phase-space explored before and after secondary decay can be accessed 
without ambiguity. This model, as well as the recently developed version for nucleon-induced 
reactions (called n-IPSE) \cite{Lac05}, has been shown to remarkably reproduce experimental 
observations.

In this work, we use the HIPSE model to address the question of the origin of the  
bimodality signal in nuclear reactions. First the experimental protocol used in refs. \cite{Tam03,Pic05} 
is recalled. It is applied to events generated with the 
HIPSE model showing that bimodality is found. Finally, we use the possibility to access 
the phase-space before the secondary decay to understand the origin of bimodality in the model.    
To do so, we have generated $10^6$ heavy ions collisions 
for the Xe+Sn system at 50 MeV/nucleon. The full impact 
parameter distribution, ranging from the grazing to the head-on 
collisions has been generated. A complete description of the model as 
well as a discussion of the hypotheses used for cluster formation 
can be found in refs. \cite{Lac04,Lac05}.  
In order to get results directly comparable to those obtained with the INDRA $4\pi$ array, 
we have filtered the events and used exactly 
the same experimental protocol (event sorting) as described in refs. \cite{Tam03,Pic05}. 
We first use a completeness criterion. 
Here, 'filtered' events, corresponding to the best detection of the QP,  
are selected (80\% of the projectile); 
this ensures an almost complete detection of the QP products.
Due to the forward detection acceptance of INDRA 
mainly semi-peripheral reactions between 5 to 8 fm are then retained.
Complete QP events are finally sorted by using the transverse energy of the light charged 
particles (Z=1,2) coming from the Quasi-Target (QT), noted $E^{QT}_{t12}$. 
By doing so, we avoid the obvious autocorrelations between the sorting observable 
(QT) and the considered system (QP). 
Note that in the experiment as well as in the simulation, the QP selection has been made 
by taking fragments with positive center-of-mass velocities \cite{Pic05}; this assumption has 
been checked with HIPSE and is indeed correct; selecting fragments coming
from the {\it true} QP source or with positive center-of mass velocities leads 
to the very same results for this analysis. 

In the study of bimodality \cite{Tam03,Pic05}, 
QT transverse energy is assumed to be indirectly related to the order parameter 
and is presented as a way of realizing a "canonical" event sorting. 
Although the transverse energy is intimately correlated to the centrality of the 
reaction, the latter assumption is, in our opinion, far from being clear because of the associated 
large mixing of impact parameters.
In the following, we will focus on the correlation 
between the largest and the second largest fragment emitted in the forward center-of-mass hemisphere. 
We then define the charge asymmetry between the 
two largest fragments $\eta_Z=(Z_1-Z_2)/(Z_1+Z_2)$ \cite{Pic05}. 
$Z_1$ and $Z_2$ are respectively the largest and the second 
largest fragment charges. 
Thus, $\eta_Z$ is close to 1 for a large asymmetry and it will be the case if an evaporation 
residue persists after de-excitation.
By variance, if $\eta_Z \approx 0$, it corresponds to a symmetric fragmentation.
\begin{figure}
\includegraphics[height=6.5cm]{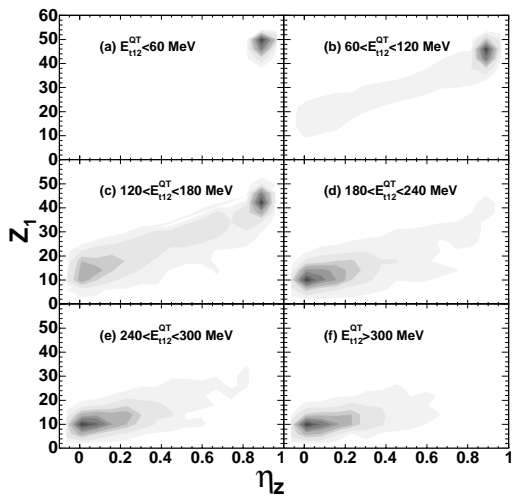}
\caption{ Correlation between the largest fragment $Z_1$ and the charge asymmetry 
$\eta_Z$ for different bins of $E_{t12}^{QT}$ (contour levels are in log. scale).}
\label{fig:epsart1}
\end{figure}
Fig. \ref{fig:epsart1} displays the correlation between $Z_1$ 
and $\eta_Z$ for different QT transverse energy intervals. 
Fig. \ref{fig:epsart1}(a) shows a single component located at
$\eta_Z \approx 1$ and $Z_1 \approx Z_{proj}$. This case 
corresponds to the evaporation residue (ER) of the projectile. 
In Fig. \ref{fig:epsart1}(b) to Fig. \ref{fig:epsart1}(f), 
we observe a different component, located this time at $\eta_Z \approx 0$ and 
$Z_1 \approx 15$; the corresponding mean fragment multiplicity is here greater than $2$, 
corresponding to the multifragmentation regime (MF). 
In Fig. \ref{fig:epsart1}(c), the 
correlation clearly exhibits both components, the asymmetric case (ER) and 
the symmetric one (MF). This coexistence has been assigned to a bimodality signal 
in the fragmenting nuclear systems \cite{Pic05}. 
Indeed, by projecting the two-dimensional distribution either on the 
$x$-axis or $y$-axis, two bumps are observed respectively 
in the distribution of $\eta_Z$ and $Z_1$ (not shown here) for the selected intermediate 
transverse energy. These results are similar to 
those obtained in the experimental case \cite{Pic05}, where a 
bimodality in $\eta_Z$ has been reported for the QP events.

\begin{figure}
\includegraphics[height=9cm]{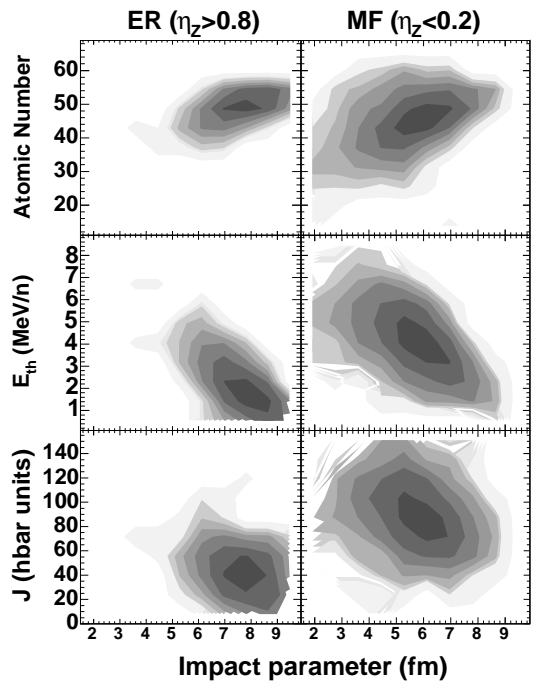}
\caption{ Correlation between the atomic number (top row), 
the thermal energy (middle) and spin (bottom)
of the QP source for events of Fig. \ref{fig:epsart1}(c). The left column corresponds to the ER case and the right to 
MF case (contour levels are in log. scale).}
\label{fig:epsart2}
\end{figure}

Let us now specify the properties of the two event classes 
observed in Fig. \ref{fig:epsart1}(b) and \ref{fig:epsart1}(c).
The charge partition may mix contributions from quasi-projectile and/or mid-rapidity emissions.
Indeed, it results from a possible {\it simultaneous} emission leading 
to a fragmented freeze-out configuration followed by a {\it sequential} decay 
of each primary fragments. In the HIPSE model both effects are taken into account 
and the initial partition before secondary decay can be easily identified. 
In particular, the number of sources contributing to the bimodality is of major 
interest. 
By tracing back the origin of clusters in HIPSE during the decay, 
we have observed that the bimodality signal displayed in Fig. \ref{fig:epsart1} 
is dominated by events where a single source is formed in the forward center-of-mass
hemisphere, this source corresponding indeed to the excited QP.  
To get a deeper insight on the origin of bimodality in HIPSE, it is 
necessary to clearly identify the phase-space explored by the QP
before de-excitation for the two event classes. Here, we concentrate 
on Fig. \ref{fig:epsart1}(c) and we will refer to ER for events with $\eta_Z>0.8$ and
MF for $\eta_Z<0.2$. 
Fig. \ref{fig:epsart2} 
presents respectively from top to bottom the correlation between the size, 
the thermal energy, the transferred angular momentum and the impact 
parameter for ER (left) and for MF (right). 
The first remarkable aspect appearing in top panels of Fig. \ref{fig:epsart2} 
is that the source sizes are not significantly different between ER ($Z_{QP} \approx 50$) and MF
case ($Z_{QP} \approx 45$) and the two distributions strongly overlap. By contrast, 
the thermal energy $E_{th}$ is much higher in the MF ($E_{th}/A \approx 4 MeV$)  
than in the ER case ($E_{th}/A \approx 1.5 MeV$). 
It is worth noticing that such a result is at variance with a geometrical scenario like 
'abrasion-ablation' models where such an increase of thermal 
energy is accompanied by a strong decrease of the QP size \cite{Gai91}. Indeed, in the HIPSE 
model, while the QP and QT are initially formed using geometrical arguments,
the abrasion picture is partially (or even completely) relaxed 
by allowing nucleon exchange and by the strong reorganisation due to Final State Interaction (FSI) \cite{Lac03}.
For the beam energy considered here ($50A.MeV$), an important exchange of particle between 
the target and projectile should also be accompanied by a large 
transfer of orbital into intrinsic angular momentum (spin). This is indeed confirmed in bottom 
part of Fig. \ref{fig:epsart2} where 
the correlation between the QP angular momentum (noted $J$) 
and the impact parameter 
is displayed. Again, we observed that while ER corresponds in average to 
$J \approx 30 \hbar$, in MF much higher angular momenta are 
obtained ($J \approx 80 \hbar$).
Consequently, the two classes of events associated to bimodality 
are issued from the de-excitation of a single-source with more or less the 
same mass but rather different initial conditions in term
of thermal energy and angular momentum. 
Note that here, the de-excitation is performed in both cases 
using the statistical sequential decay
model of ref. \cite{Dur92}. Therefore, in HIPSE, the appearance of the two contributions 
in Fig. \ref{fig:epsart1} is a direct consequence of the statistical decay accounting 
for the initial properties of the QP.   
\begin{figure}
\includegraphics[height=9cm]{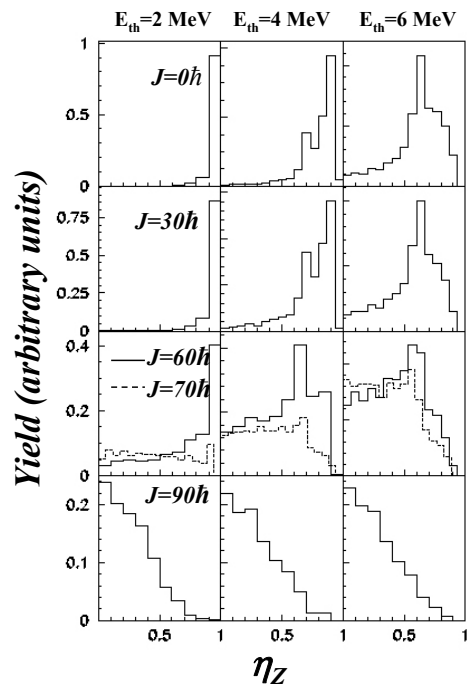}
\caption{$\eta_Z$ distributions for the statistical de-excitation of an $^{120}$Sn nucleus for different 
initial thermal energies $E_{th}$ (from left to right) and spins $J$ (from top to bottom).}
\label{fig:bim3}
\end{figure}
To go further, 
we have performed statistical simulations for a 
given nucleus ($Z=50$) by varying
the initial thermal energy and spin. 
Fig. \ref{fig:bim3} shows the corresponding $\eta_Z$ 
distributions. 
If the spin is set to zero (top row), we see that the higher the
thermal energy, the smaller the charge asymmetry is, but we never obtain 
a crossover to small values of $\eta_Z$. 
The situation is clearly different when the spin is changed. We see in Fig.  \ref{fig:bim3} 
that at low spin ($J=0\hbar$ and $30\hbar$) and high spin ($J=90\hbar$) 
only a single contribution exists respectively corresponding to high and low $\eta_Z$
values. In opposite, for spin $J=60-70\hbar$, 
we observe a sudden transition where both contributions coexist. This transition 
appears to be almost independent on the thermal energy. Therefore, 
the bimodality observed in Fig. \ref{fig:bim3} is 
related to a transition governed by the spin 
transferred in the collision, but is not of the liquid-gas type as 
concluded in \cite{Pic05}. 

The physical origin of this bimodal behaviour in the statistical decay can be inferred from
the value of the spin associated to the transition. 
Indeed, it corresponds approximately to the limit of stability of a nucleus 
with $Z \approx 50$ against prompt fission \cite{Fro96}. 
At that point, the fission barrier height becomes comparable to the energy 
of the least-bound particle and the nucleus cannot resist to the deposited spin. 
In the statistical model this corresponds to a sudden opening of decay 
channels leading to the low $\eta_Z$ contribution. 
It is worthwhile to mention that the description of such instability 
through a statistical model is certainly an approximation.
Actually, in a complete dynamical description of this instability 
we do expect that the system breaks almost at the same time as 
it is formed. In the HIPSE model, the system is assumed to be formed 
and then explore statistically accessible final configurations; it is certainly 
a too simplistic picture and calls for further theoretical developments.
\begin{figure}
\includegraphics[width=9cm]{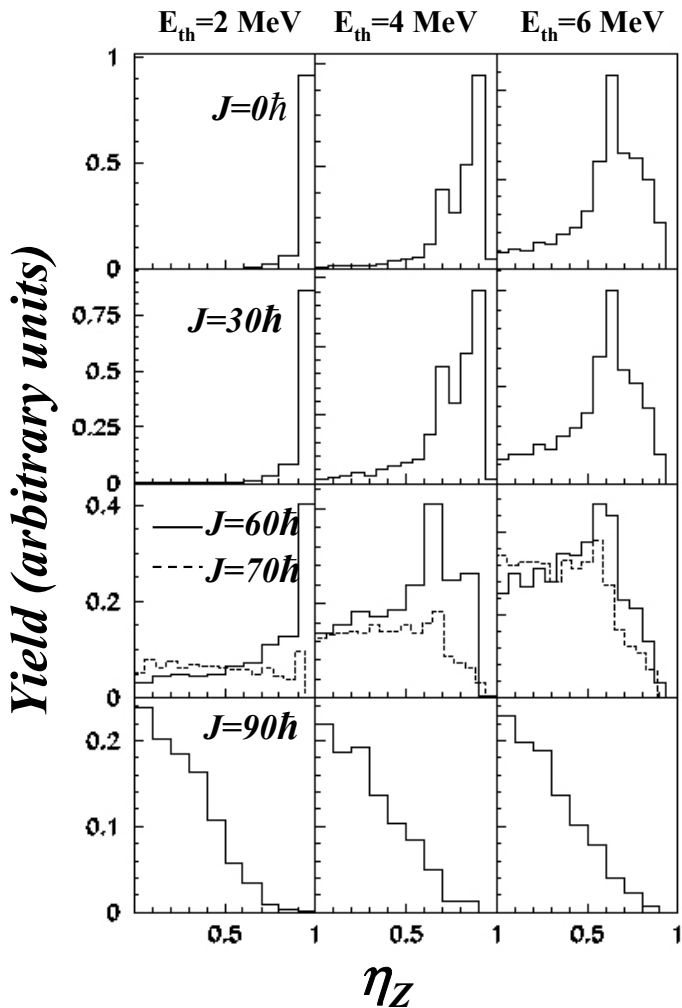}
\caption{Correlation between the largest ($Z_1$) and the 2nd largest fragment ($Z_2$) 
obtained with the standard HIPSE (left) and SMM decay (right). Both simulations 
are made with the same set of (Z,A,E*) given by the preequilibrium stage of HIPSE (contour levels are 
in log. scale). }
\label{fig:smm}
\end{figure}
In order to check our interpretation, 
we have replaced the standard HIPSE deexcitation stage by a statistical 
multifragmentation model (SMM \cite{Bond95}, using a freeze-out volume of 3 times the normal density), 
i.e. we used the HIPSE phase space before decay as an 
event-by-event input to SMM. We present in Fig. \ref{fig:smm} the correlation between 
the largest and the second largest fragment obtained with HIPSE (left panel) and the 
new decay provided by SMM (right panel). They display very similar features indicating that both scenarii 
are compatible with bimodality. Therefore, we need additional observables, for instance related to the
kinematical properties. Indeed, first investigations on experimental data indicate that 
the angular distributions of the emitted fragments are changed depending 
on the charge asymmetry selection and this could be attributed to different spin 
deposition into the decaying nucleus \cite{Steck05}. However, angular distributions need to be 
used very cautiously because of the difficulty to assess the true origin of fragments or particles
(mid-rapidity, preequilibrium, statistical emissions); it could blur the spin effect on 
the angular distributions \cite{steck01}. Such a study, comparing quantitatively HIPSE predictions to 
experimental data for QP deexcitation, will be reported in a forthcoming article.
Nevertheless, following the HIPSE scenario which has provided a good reproduction 
of a large number of experimental observations, including angular distributions for LCP and 
fragments \cite{Lac04,Van03}, we conclude that nuclear systems close or beyond their 
limit of resistance with respect to spin deposition, can be formed, leading then to a bimodal behavior.
Therefore, heavy-ion induced reactions might be a tool to study the emergence 
of shape bifurcation associated to high spin, called the Jacobi sequence \cite{Rin80}. 
This is a very interesting aspect which has not been explored in this context. If this interpretation 
is confirmed experimentally, instead of a 
liquid-gas phase transition, we may have a signature 
of the so-called Jacobi transition \cite{maj04}, which is related to a second-order 
phase transition in the continuous limit and has also its equivalent in astrophysical 
context \cite{Chr95}.

\end{document}